\documentclass[twocolumn,prl,footinbib]{revtex4}
\usepackage{graphicx}
\usepackage{epsfig}
\usepackage{dcolumn}
\usepackage{bm}
\usepackage{amsmath}
\usepackage[usenames]{color}
\newcommand{\qdc}[1]{ {\textit{\color{Red}{#1}}} }

\begin{document}
\title{Long-range spin-qubit interaction mediated by microcavity polaritons}
\author{G. F. Quinteiro Rosen$^1$, J. Fern\'{a}ndez-Rossier$^2$, and C. Piermarocchi$^1$}

\affiliation{(1) Department of Physics and Astronomy, Michigan State
University, East Lansing, Michigan 48824 USA
\\ (2) Department of Applied Physics, University of Alicante,
Alicante, 03690 Spain}

\date{\today}

\begin{abstract}
We study the optically-induced coupling between spins mediated by
polaritons in a planar micro-cavity. In the strong coupling regime,
the vacuum Rabi splitting introduces anisotropies in the spin
coupling. Moreover, due to their photon-like mass, polaritons
provide an extremely long spin coupling range. This suggests the
realization of two-qubit all-optical quantum operations within tens
of picoseconds with spins localized as far as hundreds of nanometers
apart.
\end{abstract}

\maketitle


Planar micro-cavities are semiconductor devices that confine the
electromagnetic field by means of two parallel semiconductor
mirrors. When a quantum well is placed inside a micro-cavity the
optical excitations (excitons) in the well couple to the
electromagnetic modes of the dielectric structure. In the so-called
strong-coupling regime, excitons and cavity photons give rise to new
states, cavity polaritons~\cite{weisbuch92,savona99}, which appear
in two branches separated by a vacuum Rabi splitting. In the long
wavelength limit, and for a cavity exactly at resonance with the
exciton energy, polaritons can be seen as hybrid states that are
exactly half-matter and half-light. However, even with a half-matter
character, polaritons have photon-like dispersive properties,
determined by the reduced mass of the cavity-photon/exciton system.
A photon confined in a $\lambda/2$ planar cavity has an effective
mass of $M_\gamma= \hbar n\pi/c L$ ($n$ is the refraction index and
$L$ is the length of the cavity) which is typically four to five
orders of magnitude smaller than the exciton mass. The small
polariton mass is known to affect the dynamics of optical
excitations with phonons~\cite{cassabois00,savona97}, with interface
disorder~\cite{whittaker96,savona97b}, and it suggests the
possibility of polariton Bose-Einstein condensates at room
temperature.~\cite{deng02} In this paper, we show that the small
polariton mass has also a strong effect on the mutual interaction
between spins localized in a quantum well.

Proposals for quantum computers based on spin degrees of freedom
require that individual qubits are placed close enough so to have a
significant exchange interaction between them. This exchange
interaction can be direct (i.e. induced by a controlled overlap of
the wavefunctions), or indirect when mediated by spin excitations in
a 2D electron gas~\cite{mozyrsky01} or by optical excitation across
the semiconductor
bandgap~\cite{piermarocchi02,piermarocchi04,ramon05}. In the
indirect schemes, the range of the spin coupling is related to the
mass of the mediating particles, and the coupling decreases
exponentially as a function of the distance between the spins. Here,
we show that the small polariton mass gives an extremely long range
for the spin coupling and introduces a non-exponential behavior.
This implies that spin-qubits can be located several hundreds of
nanometers apart while still retaining control on pair interaction
through the use of polaritons.

We study two localized spins in a semiconductor quantum well
embedded in a planar micro-cavity. Our numerical results consider
the case of two shallow neutral donors in GaAs (e. g. Si), but the
theory is valid with minor modifications to other impurities and
host semiconductors, as well as to charged quantum dots. Under
strong coupling conditions, optically active excitons and
intra-cavity photons combine into polaritons, while dark excitons
remain unaffected. We consider a cavity excited by an extra-cavity
continous wave (cw)  laser  with frequency below the polariton
resonance. In response to the laser, the cavity becomes polarized,
without energy absorption, and this dynamic polarization is used to
control the localized spins. Microscopically, the polarization is
described in terms of a density of virtual polaritons, proportional
to the laser intensity, that mediates an interaction between pairs
of spins.

Our first step is to use the Hopfield canonical transformation
\cite{hopfield58} which allows us to treat the coupling between
excitons and cavity-photons non-pertubatively. Due to spin-orbit and
quantum confinement we can restrict our discussion to heavy-hole
excitons only. Heavy-holes have a total spin component in the growth
direction $j_z=\pm3/2$. Optical excitations in the system are then
described by lower polaritons with $J_z=\pm 1$, upper polaritons
with $J_z=\pm 1$, and dark excitons with $J_z=\pm 2$. In the case of
donors considered here the exciton-impurity spin interaction is
given by an Heisenberg-like exchange interaction involving the
electron in the exciton and the electron in the
donor~\cite{piermarocchi04}. Therefore, if we restrict the
excitation to $\sigma_+$ circularly polarized light, optically
excited polaritons with $J_z=+1$ can only couple to states with
$J_z=+2$ by scattering with the donors. Moreover, since we are
considering microcavities in the strong coupling regime excited near
the lower polariton resonance, we will neglect the upper polariton
branch. This approximation will be further discussed below. The
final Hamiltonian consists of a term $H_0$ describing lower
polaritons with $J_z=+1$ and dark excitons with $J_z=+2$, a $H_L$
term representing the coupling of the circularly polarized cw laser
with the lower polariton at $k=0$, and an interaction $H_I$
describing polaritons and dark excitons interacting with two
localized impurities
\begin{eqnarray}\label{Eq_H}
    H_0 &=&\sum_{k} \Omega^P_{k}p^\dag_{k}p_{k}+\Omega^X_{k}b^\dag_{k}b_{k} \nonumber \\
    H_L &=& \sqrt{\mathcal{A}}\,\phi_{1s}\mathcal{V}\, r_{0} e^{i\omega_0 t}  p_{0} + h.c. \nonumber \\
    H_I &=&  \frac{J}{\mathcal{A}} \sum_{i k j k^{\prime}\lambda}e^{i({k}-k^{\prime})R_{\lambda}}
    v_{k}v_{k^{\prime}} H^{\lambda}_{ij}\nonumber~,
\end{eqnarray}
where $\{i,j\}$ refer to the type of particle, either polariton
($P$) or dark exciton ($X$) with energy $\Omega^{i}_k$ and creation
operators $p^\dag_k$ and $b^\dag_k$, respectively. The exciton
energy is $\Omega^X_k=\epsilon_0+\hbar^2 k^2/2M_X$ while the lower
polariton energy is
\begin{equation}
\Omega^P_k=\frac{\Omega^X_k+\Omega^C_k}{2}-\sqrt{\frac{(\Omega^X_k-\Omega^C_k)^2}{4}+g^2_k},
\end{equation}
where $\Omega^C_k=\hbar c/n \sqrt{k^2+(\pi/L)^2}$ is the energy of
cavity photons, and $g_k$ is the exciton-cavity photon coupling. We
will consider microcavities at resonance, i.e. satisfying the
condition $\Omega^C_0=\Omega^X_0$, with a vacuum Rabi splitting of
$2 g_0$. The index $\lambda$ identifies the two localized impurities
that we will call $A$ and $B$. $\mathcal{V}$ and $\omega_0$ are the
Rabi energy and frequency of the exciting laser, and $\phi_{1s}$ is
the excitonic enhancement in the light-matter coupling. Only the
excitonic component of the polariton at $k=0$ couples to the
external laser field. This is taken into account by the Hopfield
coefficient -which is the coefficient of proportionality between
exciton and lower branch polariton operators-
$r_0=1/\sqrt{2}$~\cite{hopfield58}. At larger $k$ the Hopfield
coefficient approaches one, since in that limit the lower polariton
becomes exciton-like. In order to take into account the finite range
of the exchange interaction we use an exciton-impurity exchange
interaction of the separable form $J v_k v_{k^\prime}$ with
$v_{k}=[1+(\Lambda \, k)^2]^{-1}$ \cite{yamaguchi54}. $R_{\lambda}$
is the position of the impurity $\lambda$, and  $\mathcal{A}$ is the
quantization area. The spin dependent operators in $H_I$ are defined
as $ H^\lambda_{XX}=\mathcal{S}^{\lambda}_{XX}b^\dag_{k^{\prime}}
b_k$, $ H^\lambda_{XP}=r_k
\mathcal{S}^{\lambda}_{XP}b^\dag_{k^\prime} p_k$,
$H^\lambda_{PX}=r^*_{k^\prime}\mathcal{S}^{\lambda}_{PX}p^\dag_{k^\prime}
b_k $,  and  $H^\lambda_{PP}=r^*_{k^\prime} r_k
\mathcal{S}^{\lambda}_{PP}p^\dag_{k^\prime} p_k$, where ${\cal
S}^\lambda_{XX} = {\cal S}^\lambda_{PP} = s^{\lambda}_z s_z$, ${\cal
S}_{PX}^\lambda = {\cal S}_{XP}^\lambda = s^{\lambda}_x s_x +
s^{\lambda}_y s_y$,  $s^\lambda$ and $s$ are electron spin operators
in the shallow donor $\lambda$ and in the polariton/exciton,
respectively. The time dependent phase in $H_L$ is eliminated by
transforming the Hamiltonian in the rotating frame at frequency
$\omega_0$.

We derive an effective Hamiltonian for the two electronic spins in
the cavity due to the presence of the external laser field. This
effective Hamiltonian can be written in terms of a level shift
operator $R(\omega_0)$~\cite{cohen98} to second order in the laser
coupling $H_L$ as
\begin{equation}
\label{lso}
 H_{eff}=\mathcal{P}R(\omega_0)\mathcal{P}=\mathcal{P}
H_L\frac{Q}{\omega_0-\mathcal{Q} H_0 \mathcal{Q} -\mathcal{Q} H_I
\mathcal{Q}}H_L\mathcal{P}
\end{equation}
where $\mathcal{Q}$ is the projector operator onto the subspace of
one excitation (polariton or dark exciton) plus the two spins, and
$\mathcal{P}$ is the projector operator onto the subspace of the two
spins without excitations.

The next step is to make an expansion of the level shift operator
operator in $H_I$ keeping terms up to second order. Our result is
therefore valid to second order both in the coupling with the
external laser, $\mathcal{V}$, and in the exchange coupling $J$.
However, since we are considering a microcavity in the strong
coupling regime, we keep all orders in the exciton-cavity coupling
by using the polariton picture. The theory can be expanded to
include higher orders in $\mathcal{V}$~\cite{fernandez04} and
$J$~\cite{piermarocchi04}. In the case of spin 1/2, those
contributions will modify the strength of the effects but will not
change the general form and the qualitative behavior of the
effective spin coupling.

The first order  in the expansion of Eq.~(\ref{lso}) in $H_I$ is an
effective magnetic field. It describes the Inverse Faraday effect of
off-resonant circularly polarized radiation with frequency close to
an optical resonance~\cite{landau84}. The Inverse Faraday effect on
impurities in semiconductors was investigated in
Refs.~\cite{combescot05,piermarocchi04}, and was observed in Mn
spins in quantum wells in Ref.~\cite{gupta01}. This term disappears
by using linearly polarized light because of time reversal symmetry.
The second order in $H_I$ plays the role of an effective interaction
between the spins. We will focus our attention to this effective
spin coupling, which is the one that can be used, for instance, to
control spin entanglement and make optically controlled quantum
gates.

The relevant terms in the level shift operator that contribute to
the spin coupling are proportional to
\begin{eqnarray*}
  H^{(2)}_{eff}  \propto H_{PX}^A G_{X}^{0}e^{i\phi}H_{XP}^B + H_{PP}^A
  G_{P}^{0}e^{i\phi}H_{PP}^B + (A \rightleftharpoons B)~,
\end{eqnarray*} where $G_{P (X) }^0$ is the bare Green's operator for the polariton
(exciton), the superscript $(2)$ indicates the second order
contribution in $J$ only, and $\phi$ is the phase arising from the
separation between impurities.

After some algebra, and using units of $\hbar =1$, we obtain the
final expression for the effective coupling as
\begin{eqnarray}\label{Eq_Heff}
 H^{(2)}_{eff} = \frac{C}{\delta^2}\left[F_{R_P} s^{A}_z s^{B}_z +  F_{R_X}(s^{A}_x s^{B}_x +
  s^{A}_y s^{B}_y) \right]
\end{eqnarray} where
$C=J^2|\mathcal{V}|^2 |\phi_{1s}|^2 |r_0|^2$, and
$\delta=\Omega^P_0-\omega_0$ is the laser-polariton detuning. The
two functions $F_{R_P}$ and $F_{R_X}$ describe the
polariton-mediated and the exciton-mediated contributions to the
spin coupling and are defined as
\begin{eqnarray*}
    F_{R_P} &=& \int_0^\infty \frac{dk}{2 \pi}\frac{r_k^2 v_k^2 k J_0(k R)}{(\omega_0-\Omega^P_k)} \nonumber \\
    F_{R_X} &=&  \int_0^\infty \frac{dk}{2 \pi}\frac{v_k^2 k J_0(k R)}{(\omega_0-\Omega^X_k)}~, \nonumber \\
\end{eqnarray*}
where  $J_0$ is the Bessel function of order zero and $R$ is the
inter-qubit separation. $F_{R_X}$ can be explicitly written in terms
of modified Bessel functions $K_0$ and $K_1$. In the large $R$ limit
the function $F_{R_X}$ has the 2D Yukawa form
\begin{equation}
F_{R_X}\sim \frac{e^{-\sqrt{2 M_X \delta} R}}{\sqrt{R}}~.
\end{equation}
Notice that the exponential behavior is characterized by a range
$\ell\sim 1/\sqrt{M_X}$. The integral $F_{R_P}$ has to be calculated
by numerical quadrature.

Eq.~(\ref{Eq_Heff}) contains the main features of the polariton
mediated spin coupling. The vacuum Rabi splitting resolves dark and
optical active excitations, making the indirect interaction mediated
by polaritons and dark excitons different in strength and form. The
overall interaction is spin-anisotropic. We can define a cut-off
wave vector $k_c$ such that for $k>k_c$ both $r_k\sim 1$ and
$\Omega^P_k \sim \Omega^X_k$. The integrands of $F_{R_P}$ and
$F_{R_X}$ coincide in that region since the polaritonic branch is
exciton-like.

\begin{figure}
  \centerline{\includegraphics[scale=1]{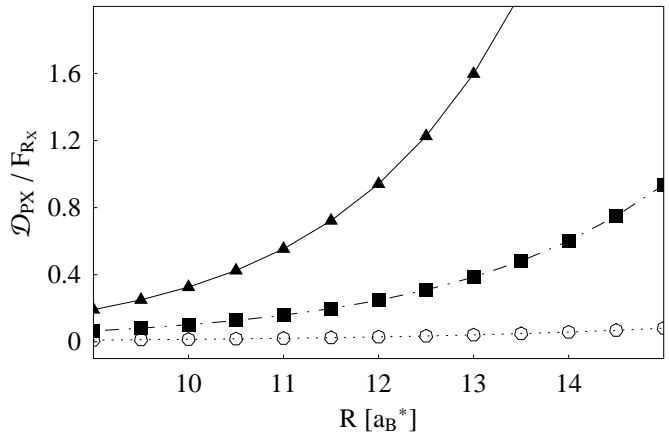}}
  \caption{Relative weight of the pure polariton to exciton contribution as
  a function of the separation between the impurities. $g_0=0.4 Ry^*$ (triangle); $g_0=0.25 Ry^*$ (square);
   $g_0=0.1 Ry^*$ (circle). In the plot we have used $\delta=0.15Ry^{*}$, $\mathcal{V}=0.16Ry^*$,
   $\Lambda=0.25 a_B^*$ and $J=21 Ry^*(a_B^*)^2$. \label{fig1}}
\end{figure}

\begin{figure}
  \centerline{\includegraphics[scale=1]{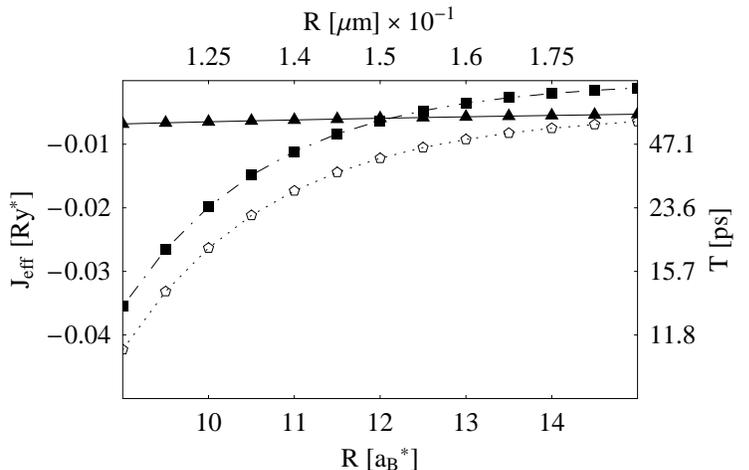}}
  \caption{Ising $J^I_{eff}= C/\delta^2{\cal D}_{PX}$ (triangle) and
Heisenberg $J^X_{eff}=C/\delta^2 F_{R_X}$ (square) contributions to
the spin coupling. For completeness, the plot shows also the
$J^P_{eff}=C/\delta^2 F_{R_P}$ (circle). The two regimes are
separated by a crossover length $R_c\simeq 12\,a_B^*$. $g_0=0.4
Ry^*$, other parameters as in Fig.~1.} \label{fig2}
\end{figure}

\begin{figure}
  \centerline{\includegraphics[scale=1]{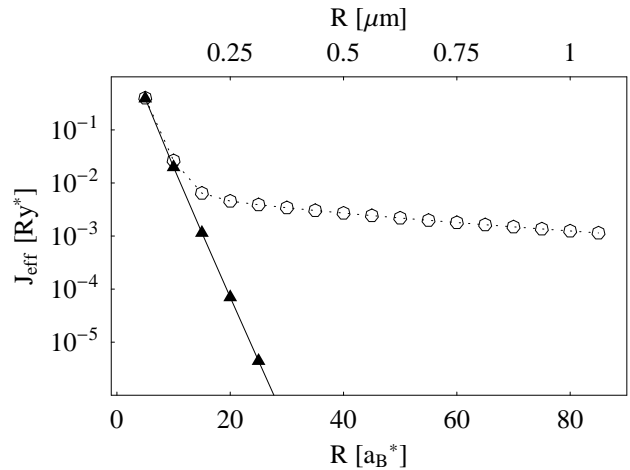}}
  \caption{Logarithmic plot of $J^X_{eff}$ (triangle) and
  $J^P_{eff}$ (circle), that shows the exponential decay and
  non-exponential behavior respectively ($g_0=0.4Ry^*$, other parameters as in Fig.~1.)
  \label{fig3}}
\end{figure}

We can then rewrite
\begin{eqnarray*}
  F_{R_P} \simeq F_{R_X} + \int_0^{k_c} (I_P-I_X) ~ dk = F_{R_X} +
  {\cal D}_{PX}
\end{eqnarray*} where $I_i$ is the integrand of either $F_{R_P}$ or
$F_{R_X}$. The term ${\cal D}_{PX}$ represents then the pure
polariton contribution, while all the excitonic effects
(dark-excitons plus polaritons at large $k$) are included in
$F_{R_X}$. The cut-off $k_c$ depends on the exciton cavity detuning
and on the strength of the exciton-cavity coupling.
Eq.~(\ref{Eq_Heff}) is rewritten as,
\begin{equation*}
  H_{eff} =\frac{C}{\delta^2} \left(
   F_{R_X} \, s^{A} \cdot s^{B}
   +  \, {\cal D}_{PX} \, s^{A}_z s^{B}_z\right)~.
\end{equation*}
Fig.~\ref{fig1} shows the relative strength of the Ising-like
polariton-mediated and the isotropic exciton-mediated contributions
for different values of the exciton-cavity coupling $g_0$ as a
function of the spin distance. Excitonic atomic units are used,
where energy is given in excitonic $Ry^*$ and lengths are in Bohr
radius a$_B^*$; for GaAs, they correspond to $1Ry^*=4.4 meV$  and
$a^*_B \sim125\AA$.
The analytical and numerical values for $F_{Rx}$, $F_{Rp}$ and
${\cal D}_{px}$ are presented in Fig.~\ref{fig2}. Notice the
existence of two distinct regions separated by a crossover distance
$R_c$. For $R<R_c$, the dominant interaction has an isotropic
Heisenberg form, while it changes to Ising-like for $R>R_c$. The
fist regime corresponds to an exciton-mediated coupling, while the
second is a pure polariton-induced effect. Polaritons, due to their
light mass, mediate a long range spin coupling interaction. However,
since they have a fixed $J_z=+1$ they cannot flip the impurity
spins, and can only induce an interaction diagonal in the impurity
spin space. The upper polariton has also $J_z=+1$ and can only
contribute to the Ising-like term. However, its contribution is
reduced by $(\delta+2 g_0)^2$ due to the vacuum Rabi splitting $2
g_0$. Fig.~\ref{fig3} shows that the strength of the exciton and
polariton induced coupling in logarithmic scale. For the exciton,
the interaction decays exponentially, while the polariton mediated
term can survive up to extremely long spin-separations and shows a
non-exponential behaviour.
The long range nature of the polariton-mediated interaction presents
important technological advantages for quantum information
implementations. Using our parameters (see Fig.~\ref{fig1}), we
predict that the strength of the interaction is $J_{eff}\doteq
1\,10^{-2}Ry^*$ for impurities separated by distances of the order
of $R_c=12\,a_b^*\simeq 150nm$. An estimate for the time needed for
an operation can be given as $T=\pi/J_{eff}\simeq 40ps$, which is
much smaller than the typical dephasing time for impurity spin
qubits. Recent measurements have reported a spin relaxation time of
the order of $\mu$s for donors in GaAs~\cite{fu05}. To our
knowledge, the spin decoherence time ($T_2$) of a single donor in
GaAs has not been measured, but is also expected to be in the $\mu$s
range. Moreover, notice that the time needed for a quantum operation
does not change considerably when we further increase the qubit
separation. Even with an inter-qubit separation of 1 micron the time
needed for one operation increases only by one order of magnitude to
about 400 ps, and is still reasonably smaller than the decoherence
time. With such a long range interaction, the realization of
electric gates to control one-qubit operations and the use of
localized magnetic field becomes feasible. The Ising-like
interaction at long separation is not a limitation for quantum gate
implementations. The polariton mediated coupling could be also used
to control the nuclear spin of the donor in a scheme similar to the
one in Ref.~\cite{mozyrsky01}. In contrast to other cavity QED-based
quantum computing implementations~\cite{imamoglu99}, the scheme
discussed here does not require 0D confined electromagnetic modes,
which is much harder to achieve experimentally. In a planar cavity
the lateral dimension is not limited by the optical wavelength,
which provides a fully scalable geometry for the qubit.

A spin coupling can also be obtained by a real polariton population
in a scheme analogue to the RKKY~\cite{ruderman54} spin coupling
mechanism. The spin interaction induced by a 0D cavity and excitons
in quantum dots has been recently investigated~\cite{chiappe05}, and
also in this 0D case the presence of a strong coupling generates
anisotropies in the spin interaction. This approach is not appealing
for quantum computing implementation since a real population of
photocarriers will add decoherence to the spin-qubit. However, it
would be interesting to expolore the dynamics of spin in the
presence of a dense polaritons population that condense in a phase
coherent state, as observed recently in II-VI
microcavities~\cite{richard05}. High-quality microcavities embedding
Mn-doped magnetic quantum wells in the strong coupling regime have
recently been realized~\cite{bahbah05}. The polariton mediated spin
coupling could be explored in these systems as a method for the
ultrafast control of the quantum well magnetization.

In conclusion, we have shown that the optical excitation of
microcavity polaritons can couple spins localized in a quantum well.
Due to the small polariton mass, the spin coupling has an extremely
long range, and at large distances is Ising-like. The interaction is
strong enough for the realization of quantum operations with spins
located as far as several hundred of nanometers apart and within a
time scale much shorter than the spin decoherence time. This
interplay of polaritons and localized spins represents a peculiar
feature of solid-state cavity QED, which has no equivalent in the
atomic case.

\acknowledgments This work was supported by NSF DMR-0312491 and by
the Donald D. Harrington Foundation at the University of Texas
Austin.

\end{document}